\documentclass[preprint]{aastex6}
\usepackage{color}
\usepackage{graphics,natbib,graphicx}
\usepackage{rotating}

\newcommand{\lapprox} {\, \lower3pt\hbox{$\sim$}\llap{\raise2pt\hbox{$<$}}\,}
\newcommand{\gapprox} {\, \lower3pt\hbox{$\sim$}\llap{\raise2pt\hbox{$>$}}\,}

\begin{document}

\title{Anomalous Cooling of Coronal Loops with Turbulent Suppression of Thermal Conduction}

\author{Nicolas H. Bian\altaffilmark{1}, Jonathan M. Watters\altaffilmark{1}, Eduard P. Kontar\altaffilmark{1}
and A. Gordon Emslie\altaffilmark{2}}

\altaffiltext{1}{School of Physics \& Astronomy, University of Glasgow, Glasgow G12 8QQ, Scotland, UK \\
(Nicolas.Bian@glasgow.ac.uk)}

\altaffiltext{2}{Department of Physics \& Astronomy, Western Kentucky University, Bowling Green, KY 42101 (emslieg@wku.edu)}

\begin{abstract}

We investigate the impact of turbulent suppression of parallel heat conduction on the cooling of post-flare coronal loops.  Depending on the value of the mean free path $\lambda_T$ associated with the turbulent scattering process, we identify four main cooling scenarios.  The overall temperature evolution, from an initial temperature in excess of $10^7$~K, is modeled in each case, highlighting the evolution of the dominant cooling mechanism throughout the cooling process.  Comparison with observed cooling times allows the value of $\lambda_T$ to be constrained, and interestingly this range corresponds to situations where collision-dominated conduction plays a very limited role, or even no role at all, in the cooling of post-flare coronal loops.

\end{abstract}

\section{INTRODUCTION}\label{introduction}

The solar corona is composed of a plasma at temperatures greater than $\sim$$10^6$~K. Since the temperature of the photosphere is only about 5800~K \citep[e.g.,][]{2003dysu.book..335P}, it follows that the corona cannot be heated by outflow of heat from the solar surface, but rather {\it in situ}. Despite decades of research, the mechanism for this heating is still the matter of debate; however, candidate mechanisms generally fall into one of two categories: heating via multiple magnetic-reconnection-driven impulsive energy releases \citep[``nanoflares,'' e.g.,][]{1988ApJ...330..474P}, or quasi-continuous (wave dissipation) heating \citep[e.g.,][]{1998ApJ...499..945L}. Both of these processes occur in confined magnetic structures or ``loops,''  and high-spatial resolution images \citep[e.g.,][]{1992PASJ...44L.181K} show that these loops appear to have a roughly constant poloidal cross-section and an approximately semi-circular toroidal shape.

As evidenced by copious soft X-ray emission, during large solar flares the plasma in coronal loops is further heated to temperatures in excess of $10^7$~K.  According to the standard flare model, this excess heating originates during an impulsive release of magnetic energy, which not only causes the plasma temperature to steadily increase but also accelerates suprathermal particles, especially electrons \citep{2011SSRv..159....3D,2011SSRv..159..107H,2011SSRv..159..301K}.  These electrons spiral around the guiding magnetic field lines, depositing their energy in the ambient atmosphere through Coulomb collisions with ambient electrons, notably in the dense layers of the solar chromosphere at the loop footpoints.  The resulting three-order-of-magnitude increase in plasma temperature \citep[e.g.,][]{1989ApJ...341.1067M} at the loop footpoints creates a strong pressure enhancement that causes the part of chromospheric plasma to be driven upward into the corona, a process typically termed ``evaporation'' \citep{1974SoPh...34..323H}.  This pressure gradient and associated upward motion persists even after the impulsive phase heating has ceased \citep{1989ApJ...341.1067M}.

According to the standard interpretation, the hot coronal plasma initially cools principally as a result of collision-dominated conduction \citep{1962pfig.book.....S} of heat toward the chromosphere \citep{2007A&A...471..271R}.  Then, as the temperature (and thus the efficiency of thermal conduction) decreases, radiation becomes the dominant cooling mechanism. However, as has been known for some time \citep[e.g.,][]{1980sfsl.work..341M}, time profiles of soft X-ray emission from flaring loops show that cooling takes far longer than the cooling times predicted from such a model.

Recently, \citet{2013ApJ...778...68R} conducted a statistical analysis of the decay-phase cooling of $72$ M- and X-class flares. A cooling profile covering the range $(16-2)$~MK is displayed in their Figure~1, and shows that on average the soft-X ray emitting plasma cools from $1.6 \times 10^{7}$~K to $10^{7}$~K in about $3$ minutes, corresponding to an average cooling rate of $\sim$$3.5 \times 10^{4}$~K~s$^{-1}$. Numerous other works (see, e.g., Figure~15 of \citet{1994SoPh..153..307C}, Figure~11 of \citet{2001SoPh..204...91A}, and Figure~5 of \citet{2006SoPh..234..273V}) support the general magnitude of this cooling time. When compared with the \citet{1995ApJ...439.1034C} cooling model (which is based on collisionally-dominated thermal conduction), the \citet{2013ApJ...778...68R} observations revealed a cooling time that was systematically greater than that predicted by the model. They attributed this to continued energy input to the corona during the decay phase, with the amounts of energy required to explain the observed cooling times lying within the range $10^{28}-5\times 10^{30}$~erg, approximately half the total energy radiated by the hot plasma.

Spatially-resolved soft X-ray observations often show localization of soft X-ray sources near the apex of flaring loops \citep[e.g.,][]{1998A&A...334.1112J,2015A&A...584A..89J}, which further suggests enhanced trapping of the hot soft-X-ray-emitting plasma. \citet{2006ApJ...638.1140J} have investigated the spatial and spectral evolution of such loop-top sources in relation to their cooling properties. They show that the instantaneous cooling rate, defined as $\dot{E}/E$ (where $E=3nk_{B}T_{e}$ is the thermal energy content), is generally two orders of magnitude lower than expected from classical thermal conduction but only slightly larger than the rate expected from radiation. They also estimated for each flare the amount of ``missing'' energy, which they interpreted either as additional energy input \citep[cf.][]{2013ApJ...778...68R} {\it or as a reduced energy loss}. Further, on the basis that this ``missing'' energy was sometimes larger than the energy input in the impulsive phase, they suggested that the latter possibility, that thermal conduction is suppressed by turbulent processes, was more likely.

Additionally, hard X-ray observations of solar flares \citep{2013A&A...551A.135S} indicate that the ratio of the number of coronally-confined electrons to the number of precipitating electrons (above $\gapprox 30$~keV) is greater than that predicted for an environment where particle transport is dominated by Coulomb collisions. \citet{2016ApJ...824...78B} have therefore proposed that scattering off turbulent magnetic fluctuations acts to reduce the efficiency of particle transport, thereby confining the high-energy electrons that produce hard X-rays \citep{2014ApJ...780..176K} to the coronal regions of the flare. \citet{2016ApJ...824...78B} point out that this turbulence will not only confine the high-energy hard-X-ray producing, electrons, but will also act to confine the lower-energy electrons that carry the conductive heat flux, thus reducing the thermal conductive heat flux below its classical \citet{1962pfig.book.....S} value and possibly accounting for the relatively long observed cooling times in accordance with the suggestion of \citet{2006ApJ...638.1140J}.

The theoretical framework for turbulent scattering in plasmas was developed some time ago \citep{1969npt..book.....S} by analogy with collisional scattering theory, with angular scattering being the predominant effect in low-frequency turbulence \citep{1966JETP...23..145R}. Scattering by the electrostatic field fluctuations of low-frequency ion-sound waves has long been invoked to explain enhanced confinement of hot electrons and reduced heat conduction during flares \citep{1979ApJ...228..592B,1980ApJ...242..799S}. However, efficient scattering of heated electrons by ion-sound turbulence requires the ions to remain cold (i.e., $T_{i}\ll T_{e}$) in order for the generated waves to overcome Landau damping.

In this work, we therefore explore the suppression of heat conduction by including scattering by low-frequency magnetic field fluctuations in the plasma. We evaluate the role of such turbulent scattering on the overall cooling of the post-flare plasma and on the transition from conduction-driven to radiation-driven cooling. Rather than including all pertinent cooling mechanisms simultaneously in a numerical treatment, we instead seek to establish temperature ranges in which each of several cooling mechanism dominates, thus yielding a piece-wise-continuous approximate analytical expression for the temperature evolution as a function of time and, more importantly, a deeper understanding of the relative roles of various cooling processes throughout the cooling period.

A significant limitation of the model is that it ignores the well-established hydrodynamic evolution of the loop during the cooling process, involving substantial transfer of mass between the chromosphere and the corona. For large downward heat fluxes, the transition region is unable to radiate the supplied energy, resulting in the deposition of thermal energy in the dense chromosphere.  The resulting 2-3 order-of-magnitude temperature enhancements create a large pressure gradient that drives an upward enthalpy flux of ``evaporating'' plasma. However, as the loop cools, the decreased heat flux becomes insufficient to sustain the radiation emitted in the now-dense transition region and hence an inverse process of {\it downward} enthalpy flux starts to occur. It has been suggested \citep{2008ApJ...682.1351K} that the enthalpy fluxes associated with both evaporating and condensing plasma are at all times in approximate balance with the excess or deficit of the heat flux relative to the transition region radiation loss rate. This basic idea has allowed the development of global ``Enthalpy-Based Thermal Evolution of Loops'' (EBTEL) models that describe the evolution of the average temperature and density in the coronal part of the loops; these models are generally in good agreement with one-dimensional hydrodynamic simulations \citep{2008ApJ...682.1351K,2012ApJ...752..161C,2012ApJ...758....5C}.
It is in principle possible to include the effects of a turbulence-controlled heat flux in EBTEL (or 1-D hydrodynamic) models.  If this heat flux is reduced sufficiently relative to its collisional value, then, for the reasons explained above, there will be a significant impact on the thermal evolution of the loop. Doing so, however, would still require a numerical treatment, which is beyond the scope of the present work (but which it is our intention to carry out in a future work). Instead we adopt a simpler approach that allows a systematic and fairly transparent quantitative analysis of the impact of turbulence on the thermodynamics of post-flare loops.

In Section~\ref{energy-terms} we provide the basic energy equation governing the heating and cooling of coronal flare plasma in a static (zero-mass-motion, constant volume and hence constant density) model, and we evaluate the order-of-magnitude values of the various cooling mechanisms involved.  In Section~\ref{conduction-radiation-phases} we provide formulae for the temperature evolution during conduction-driven and radiation-driven cooling, noting the fundamentally different evolutions that result from turbulence-dominated and collision-dominated conduction.  In Section~\ref{overall} we follow the temperature evolution of coronal plasma as it cools.  We find that there are four main ``pathways'' from an initial temperature of $1.5 \times 10^7$~K to a ``final'' temperature of $10^5$~K, depending on the value of the turbulent scattering mean free path $\lambda_T$:

\begin{itemize}

\item For very high values of $\lambda_T$ ($\gapprox 2 \times 10^8$~cm), turbulent scattering is unimportant.  Cooling thus proceeds through two main phases: collision-dominated conduction followed by radiation;

\item For somewhat lower values of $\lambda_T$ ($5 \times 10^6$~cm $\lapprox \lambda_T \lapprox 2 \times 10^8$~cm), turbulence-dominated conduction initially dominates. However, as the temperature (and with it the collisional mean free path) falls, collision-dominated conduction starts to become more important in driving conductive losses.  Cooling thus proceeds through {\it three} main phases: turbulence-dominated conduction, followed by collision-dominated conduction, and ultimately radiation;

\item For even lower values of $\lambda_T$ ($3 \times 10^5$~cm $\lapprox \lambda_T \lapprox 5 \times 10^6$~cm), the transition to radiation-dominated cooling occurs before the transition from turbulence-dominated conduction to collision-dominated conduction can occur.  Collision-dominated conduction is thus rendered unimportant, and the cooling proceeds through two phases: turbulence-dominated conduction followed by radiation;

\item For very low values of $\lambda_T \lapprox 3 \times 10^5$~cm, conduction is effectively suppressed and the cooling proceeds through a single radiative phase.

\end{itemize}

We explicitly evaluate the timescales for these various cooling phases for prescribed values of the coronal density, temperature and loop half-length, and compare with observations of actual cooling profiles in order to constrain the value of $\lambda_T$. Our conclusions are presented in Section~\ref{conclusions}.

\section{ENERGY BALANCE IN STATIC CORONAL LOOPS}\label{energy-terms}

The temperature behavior in a static coronal loop can be modeled using the usual one-dimensional energy equation

\begin{equation}\label{eq:energy}
3 n k_B \, \frac{dT}{dt} = H - L_{r}-L_{q} \,\,\, ,
\end{equation}
where $n$ (cm$^{-3}$) is the electron number density, $k_B$ is Boltzmann's constant, $H$ (erg~cm$^{-3}$~s$^{-1}$) is the volumetric heating rate, $L_{r}$ (erg~cm$^{-3}$~s$^{-1}$) is the radiative loss rate and

\begin{equation}\label{eq:lq}
L_q = - \, \frac{\partial q}{\partial s}
\end{equation}
is the conductive loss rate, with $q$ (erg~cm$^{-2}$~s$^{-1}$) being the conductive heat flux along the direction $s$ defined by the magnetic field lines.

For definiteness, we consider a coronal volume $V \sim L^{3} \simeq (2 \times 10^{9}~{\rm cm})^{3}\simeq 10^{28}$~cm$^3$, with ambient density $n \simeq 10^{10}$~cm$^{-3}$ and temperature $T \simeq 1.5 \times 10^7$~K, permeated by a magnetic field $B_{0} \simeq 300$~G, which are typical flare values \citep[e.g.][]{2012ApJ...759...71E}. The magnetic energy density $B_0^2/8\pi \simeq 3 \times 10^3$~erg~cm$^{-3}$ and the total available magnetic energy is $(B_0^2/8\pi) \, V \simeq 3 \times 10^{31}$~erg. We now consider the typical magnitudes of the terms in the energy equation~(\ref{eq:energy}):

$\bullet$ {\it Heating Rate $H$.}

If we assume that approximately one-tenth of the available magnetic energy, namely $3 \times 10^{30}$~ergs, is dissipated over a time scale $\tau \simeq 30$~s, then the average power is $P \simeq 10^{29}$~erg~s$^{-1}$ and the volumetric heating rate

\begin{equation}\label{eq:hvalue}
H \simeq 10~{\rm erg~cm}^{-3}~{\rm s}^{-1} \,\,\, .
\end{equation}

$\bullet$ {\it Radiative Loss Rate $(-L_r)$.}

For the optically thin regions of the solar atmosphere (the corona and the chromosphere where $T \gapprox 10^{4}$~K),
the radiative loss can be effectively modeled as

\begin{equation}\label{eq:lr-expression}
L_{r} = n^{2} \, \Lambda(T) \,\,\, ,
\end{equation}
where $\Lambda(T)$ (erg~cm$^3$~s$^{-1}$) is the radiative loss function \citep[e.g.,][]{1969ApJ...157.1157C,1989ApJ...338.1176C}.  A piece-wise continuous function \citep[e.g.,][]{1985ApJ...298..876A}
is commonly used to represent the radiative loss function $\Lambda(T)$, and a reasonable approximation that is useful for analytical modelling over the temperature range $10^{4}<T<10^{7}$~K is

\begin{equation}\label{eq:radt}
\Lambda(T) = \xi \, T^{-\ell} \,\,\, ,
\end{equation}
with $\xi = 1.2\times 10^{-19}$ and $\ell = 1/2$. Thus, at the assumed temperature of $T = 1.5\times 10^7$~K and density $n \simeq 10^{10}$~cm$^{-3}$, the radiative energy loss rate is

\begin{equation}\label{eq:lr-value}
L_{r} \simeq 3  \times10^{-3}~{\rm erg~cm}^{-3}~{\rm s}^{-1} \,\,\, ,
\end{equation}
some four to five orders of magnitude less than the heating rate $H$ and only weakly dependent on temperature.

$\bullet$ {\it Conductive Cooling $(-L_q)$.}

Estimating the value of the heat conduction term is somewhat more involved: it depends on the microscopic physics of the scattering of the electrons that carry the heat flux $q$.  In general, we may write

\begin{equation}\label{eq:qdtds}
q = -\kappa \, \frac{dT}{ds} \,\,\, ,
\end{equation}
where the thermal conductivity coefficient

\begin{equation}\label{kappa-general}
\kappa = \frac{2n k_B(2k_BT)^{1/2}}{m_e^{1/2}} \, \lambda \,\,\, .
\end{equation}
Here $m_e$ is the electron mass and $\lambda$ is the mean free path associated with the pertinent scattering mechanism.  For scattering by Coulomb collisions, we have \citep[see, e.g.,][]{1962pfig.book.....S}

\begin{equation}\label{lambdaei}
\lambda = \lambda_{ei} = \frac{(2k_B T)^2}{2\pi e^4 \ln \Lambda \, n} \simeq 10^4 \, \frac{T^2}{n} \,\,\, ,
\end{equation}
where $e$ (esu) is the electronic charge and $\ln\Lambda$ the Coulomb logarithm $\approx 20$.  For such a collision-dominated regime, the thermal conductivity coefficient is thus

\begin{equation}\label{eq:ks}
\kappa_S = \frac{k_B \, (2k_BT)^{5/2}}{\pi m_e^{1/2}e^4 \ln\Lambda} \equiv \alpha \, T^{5/2} \simeq 1.7 \times 10^{-6} \, T^{5/2} \,\,\, .
\end{equation}
Writing the heat flux as
\begin{equation}
q_S = -\frac{2\alpha}{7} \, \frac{dT^{7/2}}{ds},
\end{equation}
and setting $T = 1.5 \times 10^{7}$~K, we obtain the conductive loss rate in a collision-dominated regime:

\begin{equation}\label{lq-coll}
L_{qS} \simeq \frac{q_{S}}{L/2}\simeq\frac{2 \alpha}{7} \, \frac{T^{7/2}}{(L/2)^2} \simeq 5 \,\, {\rm erg~cm}^{-3}~{\rm s}^{-1} \,\,\, ,
\end{equation}
which is over three orders of magnitude larger than the radiative loss rate at this temperature. Therefore, the post-flare loop cooling is expected to be dominated by conduction.

\citet{2016ApJ...824...78B} have shown that the behavior of nonthermal electrons in certain flaring loops requires that electrons also suffer significant scattering due to processes other than Coulomb collisions. For example, interaction between the electrons and small-scale magnetic fluctuations within the flaring loop gives a turbulent mean free path

\begin{equation}\label{eq:lambdat-deltab}
\lambda_T = \lambda_B \left( \frac{\delta B_{\perp}}{B_0} \right)^{-2},
\end{equation}
where $\lambda_B$ is the magnetic correlation length and $\delta B_{\bot}$ is the magnitude of the magnetic fluctuations perpendicular to the background magnetic field, $B_0$.  In the presence of such an additional scattering process, the overall mean free path is given by adding the constituent scattering frequencies $\nu = v/\lambda$:

\begin{equation}\label{general-lambda}
\nu = \nu_{ei} + \nu_T \, ; \qquad \frac{1}{\lambda} = \frac{1}{\lambda_{\rm ei}} + \frac{1}{\lambda_ {T}} \,\,\, .
\end{equation}
Introducing the dimensionless ratio

\begin{equation}\label{eq:R}
R (T) = \frac{\lambda_{\rm ei}}{\lambda_{T}} \simeq \frac{10^4 \, T^2}{n \, \lambda_{T}} \,\,\, ,
\end{equation}
we can write $\lambda = \lambda_{ei}/(1+R)$ and hence

\begin{equation}\label{kappa-reduction}
\kappa = \frac{\kappa_{s}}{1+R} \, \qquad q = - \frac{\kappa_S}{1+R} \, \frac{dT_{e}}{ds} \,\,\, .
\end{equation}
When $R\ll 1$ ($\lambda_T \gg \lambda_{\rm ei}$), we recover the collisional \citep{1962pfig.book.....S} values of $\kappa$, $q$, and $L_q$.  However, when $R\gg1$, then the small turbulent mean free path dominates the electron transport physics.  In such a situation, the thermal conduction coefficient, the heat flux, and the loss rate $L_q$ are all reduced by a factor of $\simeq R$ compared to their Spitzer values.  In the limit $R\gg1$, the thermal conductivity coefficient is given by

\begin{equation}\label{eq:kt}
\kappa_T = \frac{2 n k_B(2k_BT)^{1/2}}{m_e^{1/2}} \, \lambda_{T} \equiv \beta \, n \, T^{1/2} \simeq 1.5 \times 10^{-10} \, \lambda_T \, n \, T^{1/2} \,\,\, .
\end{equation}
(Note that $\kappa_T$ depends more weakly on temperature than the collisional conductivity coefficient $\kappa_{S}$.) The corresponding turbulence-dominated heat flux can be written as

\begin{equation}\label{q-turbulent}
q_{T} = - \beta \, n \, T^{1/2} \, \frac{dT}{ds} = -\frac{2}{3} \, \beta \, n \, \frac{dT^{3/2}}{ds} \,\,\, ,
\end{equation}
so that

\begin{equation}\label{lq-turbulent}
L_{qT} \simeq \frac{q_T}{L/2} \simeq \frac{2}{3} \, \beta \, \frac{n T^{3/2}}{(L/2)^2} \simeq 10^{-10} \, \frac{n T^{3/2}}{(L/2)^2} \,\,\, .
\end{equation}
Inserting values $n= 10^{10}$~cm$^{-3}$, $T = 1.5 \times 10^7$~K, and $L = 2 \times 10^9$~cm, we find that

\begin{equation}\label{lq-turb}
L_{qT} \simeq 5 \times 10^{-8} \, \lambda_T \,\,\, {\rm erg~cm}^{-3}~{\rm s}^{-1} \,\,\, .
\end{equation}
This is comparable to the collision-dominated conductive heating rate~(\ref{lq-coll}) when $\lambda_T \simeq 10^8$~cm, consistent with a value $R \simeq 1$ (Equation~(\ref{eq:R})) for the parameters used.  However, it should be noted that as the plasma cools, the relatively strong $T^{5/2}$ dependence of $\kappa_S$ compared to the $T^{1/2}$ dependence of $\kappa_T$ leads to an increase in the ratio $\kappa_T/\kappa_S$. Thus, even for this value of $\lambda_T$ that lead to a turbulence-dominated conductive regime for $T \simeq 1.5 \times 10^7$~K, eventually collision-dominated conduction will dominate.

\section{CONDUCTIVE AND RADIATIVE COOLING REGIMES}\label{conduction-radiation-phases}

While cooling of the plasma is possible even when the heating term $H>0$, we will here focus on the case when heating has ceased ($H=0$), so that the energy equation can be written

\begin{equation}\label{cool}
3nk_B \, \frac{\partial T}{\partial t} = \frac{\partial}{\partial s} \left( \kappa\frac{\partial T}{\partial s} \right) - n^{2} \, \chi \, T^{-l} \,\,\, .
\end{equation}
We now explore the solution of this equation in a variety of regimes.

\subsection {Conductive Cooling Regimes}

\subsubsection{Collision-Dominated}

When conduction dominates over radiation, i.e., $L_{q}\gg L_{r}$, we have

\begin{equation}\label{eq:subk}
3nk_B \, \frac{\partial T}{\partial t} =\frac{\partial}{\partial s} \left( \kappa\frac{\partial T}{\partial s} \right) \,\,\, .
\end{equation}
In the absence of turbulent scattering (or at sufficiently large values of the turbulent mean free path, i.e., $R\ll 1$), we can use the collision-dominated \citep{1962pfig.book.....S} model $\kappa = \kappa_S$, giving

\begin{equation}\label{eq:static}
3nk_B \, \frac{\partial T}{\partial t} = \frac{2\alpha}{7} \, \frac{\partial ^{2}T^{7/2}}{\partial s^{2}} \,\,\, .
\end{equation}
Using the standard separation of variables ansatz:

\begin{equation}\label{eq:sep-var}
T = T_0 \, \theta(t) \, \phi(s) \,\,\, ,
\end{equation}
we find that the temporal part satisfies

\begin{equation}\label{eq:coll-temporal}
\frac{d \theta^{-5/2}}{dt} =\frac{1}{\tau_{cS}} \,\,\, ,
\end{equation}
which has solution

\begin{equation}\label{eq:Sstattemp}
\theta(t) = \left( 1+\frac{t}{\tau_{cS}} \right)^{-2/5} \,\,\, ,
\end{equation}
where we have introduced the characteristic cooling time

\begin{equation}\label{eq:statStau}
\tau_{cS}=\frac{21 \, nk_B L^2}{20 \, \alpha \, T_0^{5/2}} \simeq 10^{-10} \, \frac{nL^2}{T_0^{5/2}} \,\,\, .
\end{equation}
Using the values $n=10^{10}$~cm$^{-3}$, $T_0 = 1.5 \times 10^7$~K, and $L = 2 \times 10^9$~cm, the characteristic cooling time is

\begin{equation}\label{eq:tau-cool-collision}
\tau_{cS} \simeq 5~{\rm s}.
\end{equation}
(As we shall see below, however, simply establishing the initial value of the characteristic cooling time $\tau_{cS}$ does not adequately describe the cooling time profile.)

The time $t_{cool}$ it takes to cool from $T = 1.5 \times 10^7$~K to $T=10^{7}$~K is given by setting $\theta = 2/3$ in Equation~(\ref{eq:Sstattemp}), giving

\begin{equation}\label{t-cool-collision}
t_{cool} = \tau_{cS} \left [ \left ( \frac{3}{2} \right )^{5/2} - 1 \right ] \simeq 10~{\rm s} \,\,\, .
\end{equation}

Observationally, however, the time it takes for flare coronal plasma to cool from $T = 1.5 \times 10^7$~K to $T=10^{7}$~K is $t_{cool}\simeq 200$~{\rm s} \citep{2013ApJ...778...68R}. This strongly suggests that thermal conduction is suppressed relative to its collisional value, a suggestion consistent with the scenario in \citet{2016ApJ...824...78B}.  We therefore next explore conductive cooling in a model that involves turbulent scattering of the electrons that carry the conductive flux.

\subsubsection{Turbulence-Dominated Conductive Cooling}

In a turbulence-dominated regime, we substitute the expression (\ref{eq:kt}) for the turbulent conductivity into equation (\ref{eq:subk}) to obtain

\begin{equation}\label{eq:turb}
3nk_B \, \frac{\partial T}{\partial t} = \frac{2 \beta n}{3} \, \, \frac{\partial^{2} T^{3/2}}{\partial s^{2}}\,\,\, .
\end{equation}
A similar separation-of-variables analysis yields

\begin{equation}
\frac{d\theta^{-1/2}}{dt}=\frac{1}{\tau_{cT}}
\end{equation}
and hence

\begin{equation}\label{eq:Sstattemp-turb}
\theta(t) = \left( 1 + \frac{t}{\tau_{cT}} \right)^{-2} \,\,\, ,
\end{equation}
where the turbulent conductive cooling time

\begin{equation}\label{tct-expression}
\tau_{cT} \simeq \frac{9 \, k_B \, L^2}{4 \, \beta \, T_0^{1/2}} \simeq 2 \times 10^{-6} \, \frac{L^2}{\lambda_T \, T_0^{1/2}}\simeq \frac{2 \times 10^9}{\lambda_{T}} \,\,\, .
\end{equation}
Notice that the latter is independent on density. Taking a turbulent scale length $\lambda_{T}$ in the range $10^{8}-10^{7}$~cm, the characteristic cooling time is in the range

\begin{equation}\label{tau-cool-turbulence}
\tau_{cT} \simeq 20-200~{\rm s} \,\,\, ,
\end{equation}
significantly longer than the value~(\ref{eq:tau-cool-collision}) for a collision-dominated environment.  The expression for $t_{cool}$, the time it takes the plasma to cool from its initial temperature of $1.5 \times 10^7$~K to $1 \times 10^7$~K is

\begin{equation}\label{t-cool-turbulence}
t_{cool} = \tau_{cT} \left [ \left ( \frac{3}{2} \right )^{1/2} - 1 \right ] \simeq 5-50~{\rm s}
\end{equation}
for $\lambda_T = 10^8-10^7$~cm. Since $t_{cool}\sim \lambda_{T}^{-1}$, decreasing the value of the turbulent mean free path to $\lambda_{T}\simeq 5 \times 10^6$~cm gives a cooling time $t_{cool}\simeq 200$~s, which is more consistent with observations \citep{2013ApJ...778...68R}. As we shall demonstrate below, for such a value of $\lambda_T$ collision-dominated conduction plays a very limited role in the cooling of the loop.

\subsection{Radiative Cooling Regime}

When radiation dominates over conduction, i.e., $L_{r} > L_{q}$, the energy equation becomes

\begin{equation}\label{energy-radiation-dominant}
3 n k_B \frac{\partial T}{\partial t}= -n^2 \, \Lambda(T) \simeq - n^2 \chi \, T^{-\ell} \,\,\, .
\end{equation}
This can be immediately integrated to give

\begin{equation}\label{eq:radtemp}
T(t) = T_0 \left( 1- \frac{t} {\tau_r} \right)^{1/(\ell + 1)} \,\,\, ,
\end{equation}
where the radiative cooling time

\begin{equation}\label{eq:radcool}
\tau_r = \frac{3k_B}{(\ell + 1) n\chi} \, T_0^{1+l} \simeq 4 \times 10^{-16} \, \frac{T_0^{l+1}}{(\ell + 1) n\chi} \,\,\, .
\end{equation}
Taking $\chi = 1.2 \times 10^{-19}$ and $l=1/2$ results in

\begin{equation}\label{eq:radtempv}
T(t) = T_0 \left( 1-\frac{t}{\tau_r} \right)^{2/3} \, ; \qquad \tau_r \simeq 2.5 \times 10^3 \, \frac{T_0^{3/2}}{n} \simeq 1.5 \times 10^4~{\rm s} \,\,\, .
\end{equation}
To cool from $1.5 \times 10^7$~K to $1 \times 10^7$~K by this process alone would take a time

\begin{equation}\label{tcool-rad}
t_{cool} = \tau_r \left [ 1 - \left ( \frac{2}{3} \right )^{3/2} \right ] \simeq 7000~{\rm s} \,\,\, .
\end{equation}
This is much larger than the observed $t_{cool}\simeq200$~s, showing that radiation cannot be responsible for cooling
at the highest flare temperatures. In fact, by comparing the respective time scales it is easily found that radiative losses become comparable to those due to collisional conductivity at temperatures $T \lapprox 2 \times 10^{6}$~K. However, as we shall investigate further below, radiation can dominate at higher temperatures if heat conductivity is suppressed by turbulent processes.

\section{OVERALL TEMPERATURE EVOLUTION}\label{overall}

We now combine the results obtained above to describe the overall temperature evolution of a cooling loop. The extent to which each individual cooling mechanism discussed above will dominate depends on the relative values of their corresponding cooling time scales, which vary with time as the plasma cools.  As mentioned in Section~\ref{introduction}, this leads to four possible scenarios, depending on the value of the turbulent mean free path $\lambda_T$. We now proceed to establish the pertinent values of $\lambda_T$ for each case and to describe the overall temperature evolution in each situation.

\subsection{Case I: Collision-Dominated Conduction $\rightarrow$ Radiation}\label{case-I}

For sufficiently large values of the turbulence mean free path $\lambda_T$, non-collisional turbulent scattering has little effect and we recover the standard picture where cooling proceeds first by collision-dominated heat conduction followed by radiation. This case applies when $R$ is smaller than unity initially (and hence, since $R \sim T^2$, at all later times), i.e., when (Equation~\ref{eq:R})

\begin{equation}\label{rvalue-initial}
R(T_0) \simeq 10^4 \, \frac{T_0^2}{n \, \lambda_T} \lapprox 1 \,\,\, .
\end{equation}
For an initial temperature $T_0 = 1.5\times10^7$~K and density $n=10^{10}$~cm$^{-3}$, this requires

\begin{equation}\label{lambdalimitcaseI}
\lambda_T \gapprox 2 \times 10^8~{\rm cm} \,\,\, .
\end{equation}
The validity of this scenario also requires that cooling by collision-dominated conduction is more important than radiation at the initial temperature, i.e., that $\tau_r \gapprox \tau_{cS}$.  This condition (see Equations~(\ref{eq:radtempv}) and~(\ref{eq:statStau})) is

\begin{equation}\label{collisional-conduction-dominant-cooling-initial}
2.5 \times 10^3 \, \frac{T_0^{3/2}}{n} \gapprox 10^{-10} \, \frac{nL^2}{T_0^{5/2}} \,\,\, ,
\end{equation}
or

\begin{equation}\label{tstar}
T_0 \gapprox T_* \simeq 5 \times 10^{-4} \, (nL)^{1/2} \,\,\, .
\end{equation}
With $n = 10^{10}$~cm$^{-3}$ and $L = 2 \times 10^9$~cm, this gives

\begin{equation}\label{tstar-value0}
T_0 \gapprox T_* \simeq 2 \times 10^6~{\rm K} \,\,\, ,
\end{equation}
which is in fact easily satisfied. The loop therefore initially cools by collision-dominated conduction, during which the temperature behaves according to (Equations~(\ref{eq:Sstattemp}) and~(\ref{eq:statStau})):

\begin{equation}\label{t-evolution-collisional-conduction}
T(t) = T_0 \left( 1+\frac{t}{\tau_{cS}} \right)^{-2/5} \, ; \qquad \tau_{cS} \simeq 10^{-10} \, \frac{nL^2}{T_0^{5/2}} \, \simeq 5~{\rm s} \,\,\, .
\end{equation}
However, when the temperature drops to a value $T_*$, a transition from collision-dominated to radiation-dominated cooling occurs. This transition temperature is reached at a time

\begin{equation}\label{tstar-value}
t_* = \tau_{cS} \left[ \left( \frac{T_0}{T_*} \right)^{5/2}-1 \right] \simeq \tau_{cS} \left [ 5^{5/2} - 1 \right ] \simeq 750~{\rm s} \,\,\, .
\end{equation}

After $t=t_*$, cooling proceeds predominantly by radiation and the temperature evolves according to

\begin{equation}\label{tevolution-radiation-caseI}
T(t) = T_* \left( 1-\frac{(t-t_*)}{\tau_{r*}} \right)^{2/3} \, ; \qquad \tau_{r*} = 2.5 \times 10^3 \,\, \frac{T_*^{3/2}}{n} \simeq 1200~{\rm s} \,\,\, .
\end{equation}
Below temperatures $T \lapprox 10^5$~K the optically thin radiative loss function $\Lambda(T) \simeq \chi \, T^{-\ell}$ no longer holds.  Hence we set

\begin{equation}\label{final-temperature}
T_{f} = 10^5~{\rm K} \,\,\, ,
\end{equation}
as the (somewhat arbitrarily defined) ``final'' temperature. This temperature is reached at a time

\begin{equation}\label{time-to-final-temperature-overall-caseI}
t_{f} = \tau_{cS} \left[ \left( \frac{T_0}{T_*} \right)^{5/2}-1 \right]  + \tau_{r*} \, \left[ 1-\left( \frac{T_f}{T_*} \right)^{3/2} \right] \,\,\, .
\end{equation}

\begin{figure}[pht]
\centering
\includegraphics[width=0.5\linewidth]{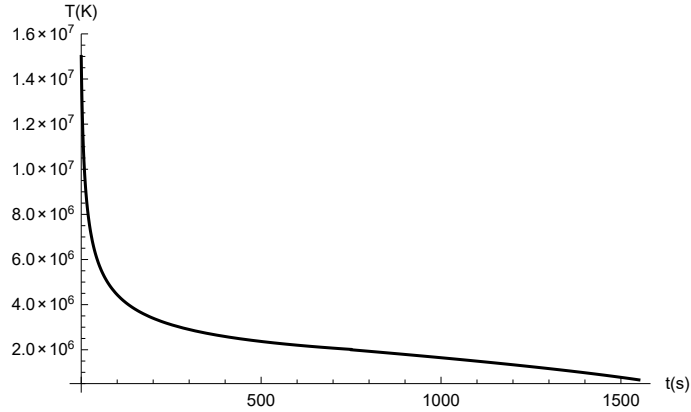}
\caption{\label{fig:CollRad} Case I: $\lambda_T = 2 \times 10^8$~cm. Cooling initially proceeds via collision-dominated conduction, transitioning to radiative cooling at $t \simeq 750$~s. Cooling to $T\simeq 10^{7}$~K proceeds by collision-dominated conduction and takes a time $t_{cool}\simeq 10$~s, a value much less than the observed cooling time $\simeq 200$~s.}
\end{figure}

The evolution of the temperature in this case is plotted in Figure~\ref{fig:CollRad}.  To summarize, cooling from $T\simeq 1.5\times 10^7$~K down to $T \simeq 2 \times 10^6$~K proceeds by collision-dominated conduction and takes a time $t_{*} \simeq 750$~s. There is then a transition to radiative cooling, which drives the temperature down to $T_f = 10^{5}$~K in a further 750~s corresponding to $t_{f} \simeq 1500$~s.

\subsection{Case II: Turbulence-Dominated Conduction $\rightarrow$ Collision-Dominated Conduction $\rightarrow$ Radiation}\label{case-II}

For values of $\lambda_{T}$ smaller than those considered in Case I, turbulent scattering is (at the initial temperature of the gas) more important than collisional scattering in determining the electron trajectories and hence turbulence-dominated conduction is more important, at least initially, than collision-dominated conduction in the evolution of the gas temperature. For such values of $\lambda_T$, we therefore expect a three-phase cooling process, starting with a turbulence-dominated conductive cooling followed as the temperature decreases by collision-dominated conductive cooling, and ending with cooling by radiation.  For turbulent scattering to initially dominate Coulomb collisions, we must have $R(T_{0})\gapprox 1$ and hence (cf. Equation~\ref{lambdalimitcaseI})

\begin{equation}\label{caseIIlambdamax}
\lambda_{T} \lapprox 2 \times 10^{8}~{\rm cm} \,\,\, .
\end{equation}
For radiation to also be negligible initially, we must have
$\tau_{r} \gapprox \tau_{cT}$, which requires (see Equations~(\ref{eq:radtempv}) and~(\ref{tct-expression})) that

\begin{equation}\label{radation-negligible-caseII}
2.5 \times 10^{3} \, \frac{T_{0}^{3/2}}{n} \gapprox 2 \times 10^{-6} \, \frac{L^{2}}{\lambda_{T} \, T_{0}^{1/2}} \,\,\, ,
\end{equation}
or

\begin{equation}\label{caseIIlambdamintent}
\lambda_{T} \gapprox 8 \times 10^{-10} \, \frac{nL^2}{T_0^{2}} \simeq 1.5 \times 10^5~{\rm cm} \,\,\, .
\end{equation}
Equation~(\ref{caseIIlambdamintent}) tentatively defines the lower limit to the value of $\lambda_T$ applicable to this case; we shall see below, however, that there is a more stringent limit on $\lambda_T$.  In the applicable regime, cooling initially proceeds via turbulence-dominated conduction, so that (see Equations~(\ref{eq:Sstattemp-turb}) and~(\ref{tct-expression}))

\begin{equation}\label{temp-turbulent-conduction}
T(t) = T_0 \left(1 + \frac{t}{\tau_{cT}} \right)^{-2} \, ; \qquad \tau_{cT} = 2 \times 10^{-6} \, \frac{L^2}{\lambda_T \, T_0^{1/2}}\simeq \frac{2 \times 10^{9}}{\lambda_T} \,\,\, .
\end{equation}
However, as the plasma cools, the collisional mean free path $\lambda_{\rm ei}$, being proportional to $T^2$ (Equation~(\ref{lambdaei})), becomes smaller.  Consequently the ratio $R(T)$ (Equation~(\ref{eq:R})), which reflects the relative importance of turbulent scattering to collisional scattering in driving the conductive heat flux) becomes smaller with time, and eventually scattering by Coulomb collisions becomes more important than collisionless pitch-angle scattering in determining the conductive cooling rate.  The temperature $T_{1}$ at which this transition occurs can be found by setting

\begin{equation}\label{r-at-t1-caseII}
R(T_1) \simeq 10^{4} \, \frac{T_{1}^{2}}{n \, \lambda _{T}} \simeq 1 \,\,\, ,
\end{equation}
giving

\begin{equation}\label{eq:Ttr}
T_{1} \simeq 10^{-2} \, (n \, \lambda_T)^{1/2}\simeq  10^{3} \, \lambda_{T}^{1/2} \,\,\, .
\end{equation}
The time $t_{1}$ at which this transition temperature is reached is, from Equations~(\ref{temp-turbulent-conduction}) and~(\ref{eq:Ttr}),

\begin{equation}\label{time-to-first-transition-caseII}
t_{1} = \tau_{cT} \left[ \left( \frac{T_0}{T_1} \right)^{1/2}-1 \right] \simeq \frac{2 \times 10^{-6} L^2}{\lambda_T \, T_1^{1/2}} \simeq \frac{2.5 \times 10^{11}}{\lambda_T^{5/4}} \,\,\, .
\end{equation}
For consistency we must check that radiative cooling remains negligible at this transition temperature. This requires that $\tau_{R} \gapprox \tau_{cS}$, which is true provided (Equation~(\ref{tstar}))

\begin{equation}\label{radiaation-negligible-at-transition-temperature-caseII}
T_1 \gapprox 5 \times 10^{-4} \, (n L)^{1/2}  \,\,\, .
\end{equation}
Comparing with Equation~(\ref{eq:Ttr}), this translates into the following condition

\begin{equation}\label{caseIIlambdamin}
\lambda_T \gapprox 2.5 \times 10^{-3} \, L \simeq 5 \times 10^6~{\rm cm} \,\,\, .
\end{equation}
Equations~(\ref{caseIIlambdamax}) and~(\ref{caseIIlambdamin}) (which is more restrictive than the tentative lower limit~(\ref{caseIIlambdamintent})) provide the respective upper and lower limits on $\lambda_T$:

\begin{equation}\label{lowerupperlimitscaseII}
5 \times 10^6 \, {\rm cm} \lapprox \, \lambda_T \, \lapprox 2 \times 10^8 \, {\rm cm}
\end{equation}
for this cooling scenario to be applicable.  The corresponding range of transition temperatures $T_1$ is, of course,

\begin{equation}\label{lowerupperlimitsT1caseII}
2\times 10^6 \, {\rm K} \lapprox \, T_1 \, \lapprox 1.5 \times 10^7 \, {\rm K} \,\,\, .
\end{equation}

During the collisional cooling phase the temperature behaves as

\begin{equation}\label{temperature-evolution-collisional-phase-caseII}
T(t) = T_{1} \left( 1 + \frac{(t-t_{1})}{\tau_{cS1}} \right)^{-2/5} \, ; \qquad \tau_{cS1} \simeq 10^{-10} \, \frac{nL^2}{T_{1}^{5/2}}
\simeq 10^{-5}\frac{L^{2}}{n^{1/4}\lambda_{T}^{5/4}}\simeq 10^{11}\lambda_{T}^{-5/4} \,\,\, .
\end{equation}
Notice that the cooling time scale $\tau_{cS1}$ in the collisional-dominated cooling regime now depends on the turbulent mean free-path $\lambda_{T}$ through the transition temperature $T_{1}$. For reasons entirely similar to those in Case I, the final transition to radiation-dominated cooling will occur at the temperature $T_* \simeq 2 \times 10^6$~K. This transition to radiative cooling is reached at a time $t_2$ obtained by solving

\begin{equation}\label{eq:T_*T_tr}
T_* = T_{1} \left( 1+\frac{(t_2-t_{1})}{\tau_{cS1}} \right)^{-2/5} \,\,\, ,
\end{equation}
giving

\begin{equation}\label{second-transition-time-caseII}
t_2 = \tau_{cT}\left[ \left( \frac{T_0}{T_1} \right)^{1/2}-1 \right] + \tau_{cS1} \left[ \left( \frac{T_1}{T_*} \right)^{5/2}-1 \right] \,\,\, .
\end{equation}

Upon reaching the temperature $T_*$, radiative cooling again finally dominates and the temperature evolves according to

\begin{equation}\label{temperature-evolution-radiative-phase-caseII}
T(t) = T_* \left( 1-\frac{(t-t_2)}{\tau_{r2}} \right)^{2/3} \, ; \qquad \tau_{r2} = 2.5 \times 10^{3} \, \frac{T_*^{3/2}}{n} \, \simeq 700~{\rm s} \,\,\, .
\end{equation}
The final temperature $T_f = 10^5$~K is reached at

\begin{equation}\label{time-at-final-temperature-overall-caseII}
t_f = \tau_{cT}\left[ \left( \frac{T_0}{T_1} \right)^{1/2}-1 \right] + \tau_{cS1} \left[ \left( \frac{T_1}{T_*} \right)^{5/2}-1 \right] + \tau_{r2} \, \left[ 1-\left( \frac{T_f}{T_*} \right)^{3/2} \right] \,\,\, .
\end{equation}

\begin{figure}\label{fig:case2}
\centering
\includegraphics[width=0.5\linewidth]{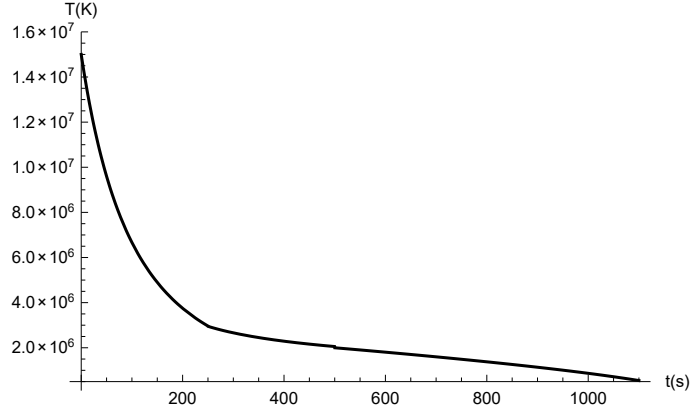}
\caption{\label{fig:3phase}Case II : $\lambda_T = 10^7$~cm. Cooling proceeds initially by turbulence-dominated thermal conduction, transitioning to collisional conduction at $t \simeq 250$~s and then to radiation at $t \simeq 500$~s.  Cooling to $T \simeq 10^{7}$~K takes a time $t_{cool} \simeq 50$~s, which is less than the observed cooling time $\simeq 200$~s.}
\end{figure}

Let us discuss a particular example. For $\lambda_{T}\simeq 10^{7}$~cm, cooling proceeds first by
turbulence-dominated conduction down to a temperature of $T_1 \simeq 3\times 10^6$~K in a time $t_{1}\simeq 250$~s.
After this, cooling proceeds by collision-dominated conduction which brings the temperature down to $T_{*}\simeq 2 \times 10^{6}$~K in a further time $\simeq 250$~s.  At this time a further transition to radiative cooling occurs and the``final'' temperature $T_{f} = 10^5$~K at the time $t_{f}\simeq 1100$~s. This case is plotted in Figure~2. We notice that a decrease in the value of $\lambda_{T}$ yields an increase in the transition time $t_{1}$ to collision-dominated cooling. This means that the duration of the collision-dominated conductive cooling regime becomes shorter with shorter $\lambda_{T}$, to the point that for $\lambda_{T}\simeq 5\times 10^{6}$~cm collision-dominated conductive cooling ends up playing little or no role at all, corresponding to a direct transition from turbulence-dominated conductive cooling to radiative cooling.

\subsection{Case III: Turbulence-Dominated Conduction $\rightarrow$ Radiation}

For values of $\lambda_T$ smaller than that given by Equation~(\ref{caseIIlambdamin}), i.e., for

\begin{equation}\label{lambdaupperlimitcaseIII}
\lambda_T \lapprox 2.5 \times 10^{-3} \, L \simeq 5 \times 10^6 \, {\rm cm} \,\,\, ,
\end{equation}
there is no intermediate collisional conductive cooling phase.  Instead, the loop will cool initially by turbulence-dominated conductive cooling and then transition directly to radiative cooling.  The temperature evolution thus proceeds in only two main phases.

Unlike in Cases I and II, where the temperature marking the transition from conductive cooling to radiative cooling is determined by equating the collisional cooling time $\tau_{cS}$ with the radiative cooling time $\tau_r$, here the transition temperature $T_{**}$ is found by equating the {\it turbulent} conductive cooling time $\tau_{cT}$ and the radiative cooling time $\tau_{r**}$ at the transition temperature, so that (Equations~(\ref{tct-expression}) and~(\ref{eq:radtempv}))

\begin{equation}\label{t**-equality-caseIII}
2 \times 10^{-6} \, \frac{L^2}{\lambda_T \, T_{**}^{1/2}}= 2.5 \times 10^3 \, \frac{T_{**}^{3/2}}{n} \,\,\, ,
\end{equation}
giving

\begin{equation}\label{eq:Tstar}
T_{**} \simeq 3 \times 10^{-5} \, \frac{n^{1/2}L}{\lambda_{T}^{1/2}} \,\,\, .
\end{equation}
The time at which this transition occurs is (Equation~(\ref{temp-turbulent-conduction}))

\begin{equation}\label{tau**caseIII}
t_{**} = \tau_{cT} \, \left[ \left( \frac{T_0}{T_{**}} \right)^{1/2} - 1 \right] \simeq 2 \times 10^{-6} \frac{L^2}{\lambda_T T_{**}^{1/2}} \simeq \frac{10^8}{\lambda_T^{3/4}}~{\rm s} \simeq 1000~{\rm s}
\end{equation}
for $\lambda_T = 5 \times 10^6$~cm.  From this time onward the loop undergoes predominantly radiative cooling until it reaches the final temperature $T_{f}$ at the time

\begin{equation}\label{eq:t**-final}
t_{f} = \tau_{cT} \, \left[ \left( \frac{T_0}{T_{**}} \right)^{1/2} -1 \right] + \tau_{r*} \,  \left[ 1-\left( \frac{T_f}{T_{**}} \right)^{3/2} \right] \,\,\, .
\end{equation}
The cooling profile for the case $\lambda_{T}\simeq 5\times 10^{6}$~cm, corresponding to a time $t_{cool}\simeq 200$~s to cool from $T\simeq 5\times 10^{7}$~K to $T\simeq 10^{7}$~K is plotted in Figure~3.

\begin{figure}\label{fig:case3}
\centering
\includegraphics[width=0.5\linewidth]{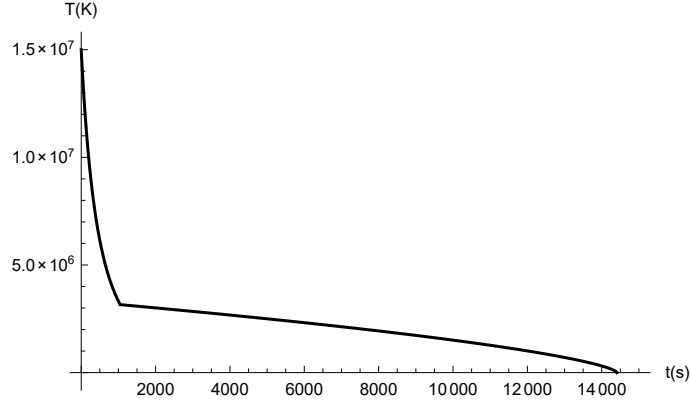}
\caption{\label{fig:TurbRad} Case III: $\lambda_{T} = 5 \times 10^{6}$~cm. Cooling initially proceeds by turbulence-dominated conduction, transitioning to radiative cooling at $t \simeq 1000$~s.  At no time does collision-dominated thermal conduction play a dominant role.  Cooling to $T \simeq 10^7$~K proceeds by turbulence-dominated conduction and takes a time $t_{cool} \simeq 200$~s, a value that is consistent with observations.}
\end{figure}

\subsection{Case IV: Radiation}

\begin{figure}
\centering
\includegraphics[width=0.5\linewidth]{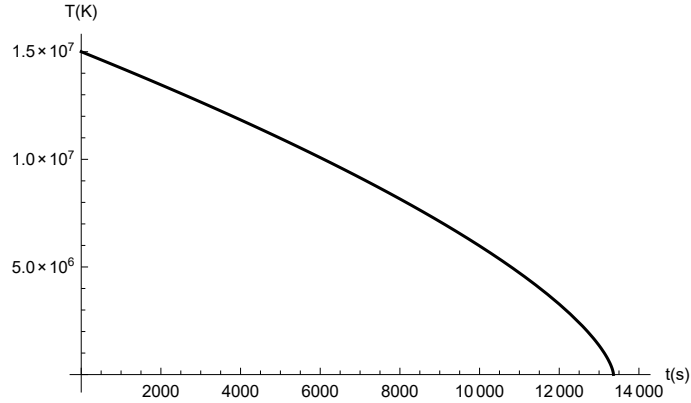}
\caption{\label{fig:radiative}Case IV: $\lambda_T = 10^6$~cm. Because conduction is highly suppressed, the cooling proceeds by radiation only, with a characteristic cooling time much larger than observed.}
\end{figure}

At the smallest values of $\lambda_{T}$ we do not expect any conductive phase at all. Indeed when $\tau_{r} \lapprox \tau_{cT}$, which requires (Equations~(\ref{eq:radtempv}) and~(\ref{tct-expression})) that

\begin{equation}\label{radiation-turbulence-equality}
2.5 \times 10^3 \, \frac{T_0^{3/2}}{n} < 2 \times 10^{-6} \, \frac{L^2}{\lambda_T T_0^{1/2}} \,\,\, ,
\end{equation}
or

\begin{equation}\label{lambdalimitcaseIV}
\lambda_{T} \lapprox 8 \times 10^{-10} \, \frac{nL^{2}}{T_{0}^2} \simeq 3 \times 10^5~{\rm cm} \,\,\, ,
\end{equation}
thermal conduction is fully suppressed and cooling proceeds primarily via radiation only. The resulting cooling profile is given by

\begin{equation}\label{tempearure-cooling-profile-radiation-caseIII}
T(t) = T_0 \left( 1-\frac{t}{\tau_{r}} \right)^{2/3} \, ; \qquad \tau_{r} = 2.5 \times 10^3 \, \frac{T_0^{3/2}}{n} \simeq 13000~{\rm s} \,\,\, ,
\end{equation}
and is plotted in Figure~\ref{fig:radiative}.

\section{SUMMARY AND CONCLUSIONS}\label{conclusions}

We have investigated the cooling of a typical post-flare coronal loop of length $L=2\times 10^9$~cm with initial temperature $T=1.5\times 10^7$~K and plasma density $n=10^{10}$~cm$^{-3}$. By varying the turbulent mean free path $\lambda_T$ we were able to identify and characterize four different cooling scenarios, summarized in Table~\ref{tableI} (see also Equations~(\ref{lambdalimitcaseI}), (\ref{lowerupperlimitscaseII}), (\ref{lambdaupperlimitcaseIII}) and~(\ref{lambdalimitcaseIV})):

\begin{table}
\caption{\label{tableI} Four different cooling scenarios.}
\begin{center}
    \begin{tabular}{ l  l  l }
    \hline
    Case  & $\lambda_T$~(cm) & Cooling Sequence \\
    \hline
    $I$ & $ > 2 \times 10^8$ & Collisional Conduction $\rightarrow$ Radiation \\ 
    $II$ & $5 \times 10^6 - 2 \times 10^8$ & Turbulent Conduction $\rightarrow$ Collisional Conduction $\rightarrow$ Radiation  \\
    $III$ & $3 \times 10^5 - 5 \times 10^6$ & Turbulent Conduction $\rightarrow$ Radiation \\
    $IV$ & $< 3 \times 10^5 $ & Radiation \\
    \hline
    \end{tabular}
\end{center}
\end{table}

Comparison of the cooling profiles with observations yields a very interesting result.  Typically, it is observed \citep[e.g.,][]{2013ApJ...778...68R} that flaring coronal loops cool from $1.5 \times 10^7$~K to $10^7$~K in about $200$~s. For the assumed loop length $L = 2 \times 10^9$~cm, this requires a value of $\lambda_T \simeq 5 \times 10^6$~cm; similar values of $\lambda_T$ result from other plausible values of $L$. This value for the turbulent mean free path falls precisely into the transition between case~II and case~III above, where collision-dominated conduction plays a very limited role, or even no role at all, in the cooling of post-flare coronal loops.  This result has very significant implications both for the modeling of cooling post-flare loops and for our understanding of the physical conditions that exist within them.

\acknowledgments This work is partially supported by a STFC consolidated grant. AGE was supported by grant NNX10AT78G from NASA's Goddard Space Flight Center.

\bibliographystyle{apj}
\bibliography{anomalous_cooling}

\end{document}